\documentclass[twocolumn,showpacs,superscriptaddress,nofootinbib]{revtex4-1}
\usepackage{graphicx}
\usepackage{subfigure}
\usepackage{epsfig}
\usepackage{epstopdf}
\usepackage{color}
\usepackage{amsmath}
\usepackage[utf8]{inputenc}

\newcommand{\be}{\begin{equation}}
\newcommand{\ee}{\end{equation}}
\newcommand{\ba}{\begin{eqnarray}}
\newcommand{\ea}{\end{eqnarray}}
\newcommand{\beq}{\begin{equation}}
\newcommand{\eeq}{\end{equation}}
\newcommand{\beqa}{\begin{eqnarray}}
\newcommand{\eeqa}{\end{eqnarray}}

\usepackage[normalem]{ulem}

\begin{document}

\title{Thermodynamic Instabilities of Generalized Exotic BTZ Black Holes}

\author{Wan Cong}
\email[]{wcong@uwaterloo.ca}
\author{Robert B. Mann}
\email[]{rbmann@uwaterloo.ca}
\affiliation{Department of Physics and Astronomy, University of Waterloo, Waterloo, Ontario,  Canada, N2L 3G1}
\affiliation{Perimeter Institute for Theoretical Physics, 31 Caroline St. N., Waterloo, Ontario, Canada, N2L 2Y5}

\date{\today}

\begin{abstract}
We examine the  conjecture  that black holes  violating the reverse isoperimetric inequality have negative specific heat at constant volume $C_V$ \cite{Johnson:2019mdp}. We test this conjecture on the family of generalized exotic Ba\~nados, Teitelbiom and Zanelli (BTZ) black holes and find that $C_V$ can be positive even when the reverse isoperimetric inequality is violated, providing a counter example to the conjecture.
However in all cases where $C_V$ is positive, the specific heat at constant pressure $C_P$ is negative,
indicating that generalized exotic black holes are thermodynamically unstable, suggesting that a broader version of the conjecture might hold.
\end{abstract}

\maketitle

In the framework of extended black hole thermodynamics \cite{Kubiznak:2016qmn,Cvetic:2010jb,Dolan:2010ha,Kastor2009,Caldarelli:1999xj}, the cosmological constant $\Lambda$ is interpreted as a pressure variable, $P=-\frac{\Lambda}{8 \pi}$, whose variation leads to the extended first law, $dM = TdS + VdP$. Identifying the black hole mass $M$ as the enthalpy of the black hole, this looks like the usual first law of thermodynamics for systems with volume. The volume term $V$ in the extended first law is the conjugate potential to $P$, with $V = \frac{\partial M}{\partial P}|_{S}$. However this ``volume'' is not simply the geometric volume of the black hole except in simple cases such as the Anti-de Sitter (AdS) Schwarzchild black hole. Indeed, one need not even have a black hole present to define this volume; a single cosmological horizon is sufficient \cite{Mbarek:2016mep}.

It has remained an ongoing puzzle to  better understand the physical interpretation of this volume term.
An early attempt to do so \cite{Cvetic:2010jb} led to the conjecture that  asymptotically AdS black holes satisfy  the {\it reverse isoperimetric inequality}, 
\begin{equation}\label{reviso}
    \mathcal{R} = \bigg(\frac{(d-1)V}{\omega_{d-2}} \bigg)^{\frac{1}{d-1}}\Big(\frac{\omega_{d-2}}{A} \Big)^{\frac{1}{d-2}}\geq 1
\end{equation}
for a $d-$dimensional spacetime. In this expression, $\omega_{d-2}$ is the dimensionless surface area of the unit ball in $d-1$ dimensions and $A$ is the surface area of the black hole horizon.  The conjecture was motivated by the observation at the time that all known black holes seemed to respect the
inequality \eqref{reviso}. When the equality is saturated, a black hole of a given thermodynamic volues has attained its maximal possible entropy. 

Since then, the isoperimetric ratio $\mathcal{R}$ of various black hole families have been studied, and some were found to violate this inequality \cite{Hennigar:2014cfa,Klemm:2014rda,Hennigar:2015cja,Brenna:2015pqa,Noorbakhsh:2016faj,Noorbakhsh:2017tbp,Feng:2017jub,Noorbakhsh:2017nde,Mann:2018jzt}. 
Such black holes have come to be called {\it super-entropic}, since they have more entropy that
the relation \eqref{reviso} would admit.  

What is still missing however is a physical interpretation of such violations. A recent proposal 
\cite{Johnson:2019mdp} hopes to remedy this shortfall by positing that  violation of the reverse isoperimetric inequality implies a new type of thermodynamic instability, namely that the specific heat at constant volume, $C_V$, of the black hole will be negative. This was shown to be the case analytically for the $(2+1)$-dimensional charged Ba\~nados, Teitelbiom and Zanelli  BTZ black holes \cite{BanadosEtal:1992}
as well as numerically for the ultra spinning $d$-dimensional Kerr black hole \cite{Hennigar:2015cja}.
Systems with negative specific heat are unstable -- an initial loss of energy is accompanied by an increase in temperature, leading to faster energy loss and a further increase in temperature. The system thus accelerates through this downward spiral rather than settling down to a nearby equilibrium state. In the case of black holes, this loss of energy can for example be due to Hawking Radiation. Traditionally, such instabilities were studied by looking at the sign of $C_P$, the specific heat at constant pressure. With the introduction of the volume term in extended thermodynamics it is natural to also look at $C_V$, which may harbour interesting physics \cite{Johnson:2019mdp,Johnson:2019vqf}. 

Here we test the conjecture of \cite{Johnson:2019mdp} using the family of ``generalized exotic BTZ black holes'', a term coined in \cite{Frassino:2019fgr}.  Such black holes violate the inequality \eqref{reviso}
and so are superentropic.  The relative simplicity of the spacetime makes them ideal candidates for
studying their specific volumes and so testing the new instability conjecture.  We find that 
generalized exotic BTZ black holes do not always have $C_V < 0$, and so provide a counterexample
to the conjecture.  However we also find that whenever $C_V > 0$ that $C_P < 0$, indicating such black holes are in general thermodynamically unstable. This suggests   a broader version of the instability conjecture:  all super-entropic black holes are thermodynamically unstable. 

A study of the thermodynamics of BTZ black holes 
\cite{Frassino:2015oca} indicated that charged BTZ black holes were super-entropic.  Generalized
BTZ black holes are also super-entropic, but demonstration of this is somewhat more subtle.
These objects are $(2+1)$-dimensional black holes whose metric is given by the usual rotating BTZ metric
\begin{equation}
    ds^2 = -f(r)dt^2+\frac{1}{f(r)}dr^2+r^2(d\phi-\frac{4 j}{r^2}dt)^2\,
\end{equation}
with
\begin{equation}
    f(r) = -8m+\frac{r^2}{\ell^2}+\frac{16j^2}{r^2}\,,
\end{equation}
where $\ell$ is the AdS radius. Here and below, we work with units in which $\hbar = G = c = k_B = 1$. 
The roots of the function $f(r)$ give the inner and outer black hole horizons, $r_{\pm} = 2\sqrt{\ell(\ell m\pm\sqrt{\ell^2m^2-j^2})}$. This metric is a solution to both the usual 3D Einstein action with a negative cosmological constant, as well as the parity-odd Einstein action \cite{Witten:1988hc}.  The expressions
for the conserved mass $M$ and angular momentum $J$ as functions of the parameters $m$ and $j$ depend on the choice of action \cite{Carlip:1994hq}.
In the former case $m$ and $j$
are respectively the mass and angular momentum of the black hole, while for the latter action we have the ``exotic'' case, with \cite{Townsend:2013ela}
\begin{equation}
\label{eq: standard}
    M  = j/\ell\,,\quad J = \ell m.
\end{equation}
This reversed role of the $m$ and $j$ parameters has also been found in other gravity models \cite{Carlip:1994hq,Carlip:1991zk,PhysRevD.57.1068,Ba_ados_1998}. 

If one takes the gravitation action to be  $I = \alpha I_{EH}+\gamma I_{GCS}$, which is a linear combination of the Einstein-Hilbert (EH) action and the gravitating Chern-Simons (GCS) action
\cite{Frassino:2019fgr,Solodukhin:2005ah}, 
then one obtains the generalized exotic BTZ black holes, for which
\begin{align}
\label{eq: exotic}
    M = \alpha m +\gamma \frac{j}{\ell}\,,\\
    J = \alpha j+\gamma \ell m\,
\end{align}
are the conserved mass and angular momentum  \cite{Frassino:2019fgr}, with $\alpha\in(0,1)$ and $\gamma = 1-\alpha$.  When $\alpha = 1$, we have the standard BTZ black hole while $\alpha = 0$ corresponds to the exotic BTZ black hole.

The thermodynamic variables of the generalized exotic BTZ black holes are \cite{Frassino:2019fgr}
\begin{align}
    M& = \frac{\alpha\big(r_-^2+r^2_+\big)}{8 \ell^2}+\frac{\gamma r_-r_+}{4\ell^2}\,,\\
    J& = \frac{\alpha r_-r_+}{4\ell}+\frac{\gamma\big(r_-^2+r^2_+\big)}{8\ell}\label{eq: J}\,,\\
    T& = \frac{r_+^2-r^2_-}{2\pi\ell^2r_+} \,,\quad \Omega = \frac{r_-}{r_+\ell}\label{eq: T}\,,\\
    S&= \frac{1}{2} (\pi\alpha r_+ +\pi\gamma r_-)\label{eq: S}\,,\\
    V&= \alpha \pi r_+^2+\gamma \pi r_-^2\bigg(\frac{3r_+}{2r_-}-\frac{r_-}{2r_+}\bigg)\label{eq: V}\,.
\end{align}
They satisfy the first law, $dM = TdS+VdP+\Omega dJ$ as well as the Smarr relation, $0=TS-2PV+\Omega J$. The ratio $\mathcal{R}$ can be readily computed to be \cite{Frassino:2019fgr}
\begin{equation}
    \mathcal{R} = \frac{1}{2}\sqrt{4\alpha-2\gamma \frac{r_-^3}{r_+^3}+6\gamma\frac{r_-}{r_+}}\,.
\end{equation}
The reverse isoperimetric inequality is saturated for standard BTZ black holes ($\alpha = 1$) with $\mathcal{R}=1$. For all other $\alpha$ values, $\mathcal{R}\leq 1$ with equality if and only if $r_-=r_+$, i.e., the case of extremal black holes. Hence, the reverse isoperimetric inequality is violated for all generalized exotic BTZ black holes except for the standard BTZ and extremal cases. We thus expect these black holes to have negative specific heat according to the conjecture in \cite{Johnson:2019mdp}.

We begin our study of thermodynamic stability by looking at the $C_P = T\frac{\partial S}{\partial T}|_{P,J}$ of generalized exotic BTZ black holes. Analytic expression of $C_P(r_+,P,J)$ can be obtained by first solving for $r_m$ in terms of $(J, P,r_+)$ using eq.~\ref{eq: J}, and then substituting this into the expressions for temperature, eq. ~\ref{eq: T}, and entropy, eq.~\ref{eq: S}. We can then use the relation $\frac{\partial S}{\partial T}|_{P,J} =\frac{\partial S}{\partial r_+}|_{P,J}\frac{\partial r_+}{\partial T}|_{P,J}$ to find $C_P$. 

\begin{figure}
    \centering
    \includegraphics[scale = 0.8]{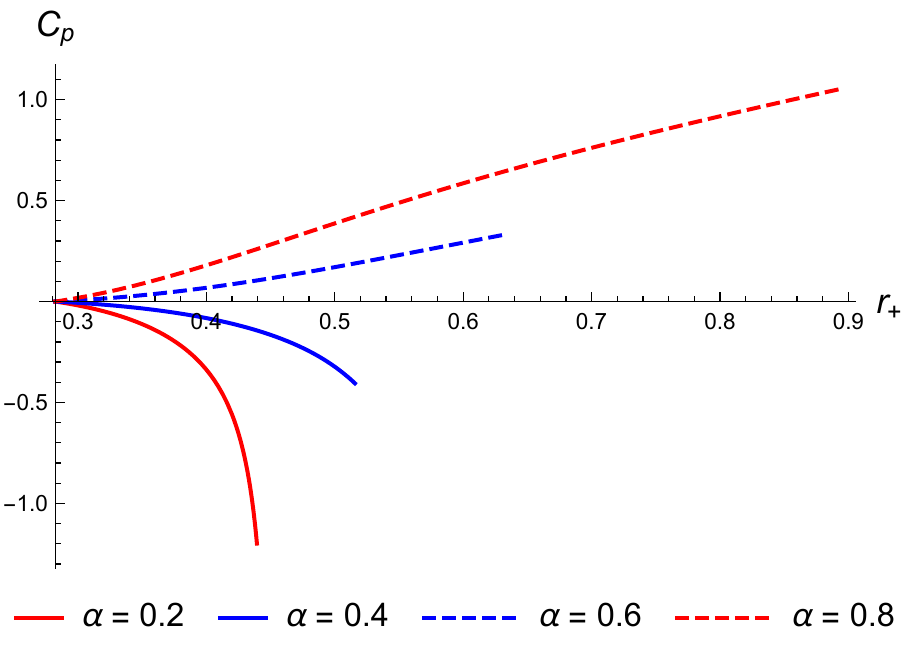}
    \caption{Plot of $C_P$ against $r_+$ for various values of $\alpha$ at $P=1$ and $J=0.1$. The range of $r_+$ in each of these plots are restricted to $\bigg(\frac{2 J^2}{\pi P} \bigg)^{\frac{1}{4}}\leq r_+ \leq \bigg(\frac{8 J^2}{\pi P(1-\alpha)^2}\bigg)^{\frac{1}{4}}$ to ensure $0\leq r_m\leq r_+$.}
    \label{fig: cp}
\end{figure}

The result is shown in fig.~\ref{fig: cp} (see also \cite{Frassino:2019fgr}). The range of $r_+$ on this figure is chosen to ensure $0\leq r_m\leq r_+$, which translates into the condition 
\begin{equation}
r_E \equiv \bigg(\frac{2 J^2}{\pi P} \bigg)^{\frac{1}{4}}\leq r_+ \leq \bigg(\frac{8 J^2}{\pi P(1-\alpha)^2}\bigg)^{\frac{1}{4}}\; .
\end{equation} 
 The black hole is extremal if $r_+=r_-=r_E$. For ``more exotic'' black holes i.e., $\alpha < 1/2$ (solid curves) $C_P$ is negative except for extremal cases where it is zero while for more standard black holes ($\alpha > 1/2$) $C_P$ is positive except again at $r_+=r_E$ where it is zero. Already, we are seeing that more exotic black holes are unstable according to the sign of $C_P$.

We now move on to look at $C_V$ for these black holes. To do this, we make use of the expression \cite{Johnson:2019mdp}
\begin{equation}
    \label{eq: cvEx}
    C_V = C_P - TV\alpha_T^2\kappa_T\,,
\end{equation}
where $\alpha_p\equiv V^{-1} \partial V/\partial T|_{P,J}$ and $\kappa_T \equiv -V\partial P/\partial V|_{T,J}$. Each of these  expressions can be evaluated in a similar manner to $C_P$.
\begin{figure}
    \centering
    \includegraphics[scale = 0.6]{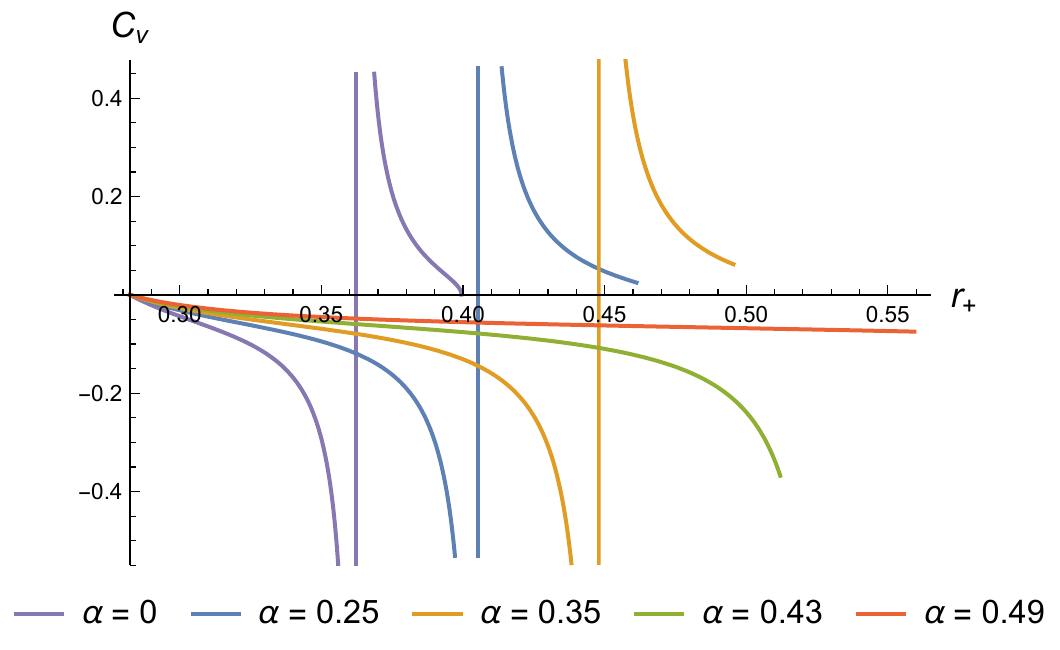}
    \includegraphics[scale = 0.6]{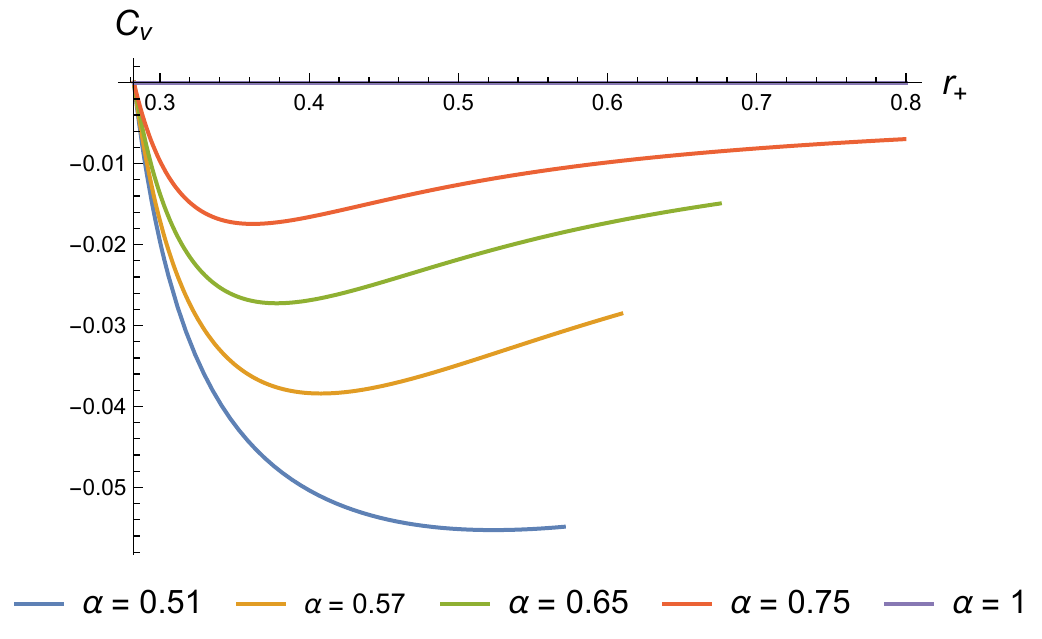}
    \caption{$C_V$ of generalized exotic black holes. Top: $\alpha < 1/2$, bottom: $\alpha > 1/2$. In all cases, $P=1, J=0.1$.}
    \label{fig: cv}
\end{figure}

The result is shown in fig.~\ref{fig: cv}. First, notice that $C_V$ of $1/2 < \alpha<1$ (bottom figure) black holes are negative, in agreement with the conjecture in \cite{Johnson:2019mdp}.  When $\alpha = 1$, $C_V = 0$, as expected for the standard BTZ black hole. 
Recall that the $C_P$ of these black holes are non-negative. This thus illustrates how the thermodynamic stability of black holes depends on whether we are keeping $P$ or $V$ fixed. 

More interesting is the $\alpha<1/2$ (top figure) case. When $\alpha$ is close to $1/2$, the $C_V$ remains negative, as in the red and green curve in the figure. These curves terminate at the extremal limit $r_+ = r_-$. However, as the black hole becomes more exotic with decreasing $\alpha$, we see that $C_V$ can become positive at large enough $r_+$ as is the case for the blue, yellow, and purple ($\alpha = 0$) curves.  The divergences in $C_V$ occur whenever $\partial V /\partial P|_{T,J} = 0$.

Summarizing, we have shown that generalized exotic BTZ black holes with a sufficiently small $\alpha$ parameter are counter-examples to the conjecture in \cite{Johnson:2019mdp}: they violate the reverse isoperimetric inequality but possess positive $C_V$.  However we note that an 
alternate (weaker) interpretation\footnote{We thank Clifford Johnson for pointing this out.} of the conjecture might be   that any \textit{connected branch} of black holes that is super-entropic will have $C_v<0$ for at least some part of the branch and therefore the entire branch is thermodynamically unstable.  While our results are consistent with this (weaker) interpretation, we regard it to be 
less precise since the meaning ``connected branch'' of black holes is somewhat ambiguous\footnote{For instance in the charged Schwarzschild-AdS solution, a single temperature can admit two black hole solutions at the same pressure and charge. These belong respectively to  ``small black hole" and ``large black hole" solution branches (see for example \cite{Kubiznak:2012wp} for more details). Such distinctions do not occur in the generalized exotic BTZ case \cite{Frassino:2019fgr}. However the 
divergence in $C_v$  can be argued to separate the class of exotic solutions into different branches.}. 

We have also observed that $C_P < 0$ whenever  $C_V> 0$.  We therefore posit a further broader conjecture that {\it black holes violating the reverse isoperimetric inequality will be thermodynamically unstable}. It will be interesting to see if this broader conjecture can either be proven or if a counterexample can be found.
 
 \section*{Acknowledgments}
 \label{sc:acknowledgements}
 We would like to thank Antonia Micol Frassino and Jonas Mureika for helpful comments, and Clifford
 Johnson for helpful correspondence. This work was supported in part by the Natural Sciences and Engineering Research Council of Canada.

\bibliography{ExoticCv-v6}

\end{document}